# Thermal Diffusivities of Functionalized Pentacene Semiconductors

H. Zhang[a], Y. Yao[a], Marcia M. Payne[b], J.E. Anthony[b], and J.W. Brill[a]

[a]Department of Physics and Astronomy, University of Kentucky,
Lexington, KY 40506-0055, USA
[b]Department of Chemistry, University of Kentucky, Lexington, KY 40506-0055, USA

**Abstract:** We have measured the interlayer and in-plane (needle axis) thermal diffusivity of 6,13-bis(triisopropylsilylethynyl) pentacene (TIPS-Pn).  The needle axis value is comparable to the phonon thermal diffusivities of quasi-one dimensional organic metals with excellent π-orbital overlap, and its value suggests that a significant fraction of heat is carried by optical phonons.  Furthermore, the interlayer (**c**-axis) thermal diffusivity is at least an order of magnitude larger, and this unusual anisotropy implies very strong dispersion of optical modes in the interlayer direction, presumably due to interactions between the silyl-containing side groups.  Similar values for both in-plane and interlayer diffusivities have been observed for several other functionalized pentacene semiconductors with related structures.

Organic semiconducting materials are being studied for a variety of room temperature applications, especially where low-cost conformal materials that can be coated on complex structures are desired. Although most of the emphasis has been on studies of electronic and optical properties, it is also important to know the thermal conductivities ($\kappa$) for different functions. For example, for micron-channel thin-film transistors, e.g. to drive pixels in flexible displays, one requires a relatively large thermal conductivity to minimize Joule heating; to keep heating less than ~ 10 degrees, one needs $\kappa > \kappa_0$, where $\kappa_0 \sim 10$ mW/cm·K [1]. On the other hand, low thermal conductivities, e.g. $\kappa < \kappa_0$, are needed for applications as room-temperature thermoelectric power generators [2].

Because small molecule, crystalline organic semiconductors can have high charge carrier mobilities and offer versatile processing possibilities, they can be advantageous for some applications. For thermoelectric applications, for which materials must be chemically doped to have sufficient conductivities, this will require substitutional doping with structurally similar molecules so as not to overly increase scattering due to disorder [2,3]. If this can be achieved, the high mobility may allow very low doping levels to be used [4], increasing the Seebeck coefficient without significantly increasing the thermal conductivity [2]. For example, layered crystals of rubrene has been found to have in-plane [5] and interplane [6] thermal conductivities comparable to that of disordered polymers [7,8].

In this regard, 6,13-bis(triisopropylsilylethynyl) pentacene (TIPS-Pn) [9-11], for which self-assembled crystalline films can be cast from solution and for which hole mobilities $\mu > 10$ cm$^2$/V·s [12] have been achieved, is a promising material. In this work, we discuss measurements of the room temperature thermal conductivities of TIPS-Pn [9-12] and several other pentacene based materials with related structures, shown in Figure 1 [13-15]. Because these crystals are undoped, the electronic density is negligible and $\kappa$ is overwhelmingly due to phonons [16]. All of these materials form layered crystals, with the pentacene (or substituted pentacene) backbones lying approximately in the **ab**-plane (where one hopes for good $\pi$-orbital overlap between molecules [9]), and the silyl or germyl side groups extending between layers [9-15]. For example the "brick-layer" **ab**-plane structure [9,11] of TIPS-Pn is shown in Figure 1g. The interlayer (**c**-axis) and in-plane phonon thermal conductivities are therefore expected to be different and we measured them separately. Because the crystals are small (in-plane dimensions typically 0.5-10 mm and interlayer direction less than 0.6 mm), we used ac-calorimetry techniques [6,17,18], which yield the thermal diffusivity, $D \equiv \kappa/c\rho$, where c is the specific heat and $\rho$ the mass density, rather than conventional techniques.

The interlayer measurement technique is described in detail in Ref. [6]. Visible/NIR light, chopped at frequency $\omega$, illuminates and heats the top surface of the opaque sample, giving a typical dc-temperature rise of the sample of a few degrees for a sample in vacuum. The temperature oscillations on the bottom surface are measured with a flattened, 25 μm diameter chromel-constantan thermocouple glued to the surface with silver paint [6,19]. The oscillating thermocouple signal, $V_\omega$, is measured with a 2-phase lock-in amplifier. If the chopping period is much less than the time constant with which the sample comes to equilibrium with the bath ($\tau_1 \gg 1$ s for our samples), the temperature oscillations of the bottom of the sample are given by

$$T_\omega = 2P_0 A \Phi(\omega)/(\pi\omega C) \quad (1a)$$

with $\Phi(\omega) \approx [1 + (\omega\tau_{meas})^2]^{-1/2} \quad (1b)$.

Here $P_0$ is the intensity of the light, A the area of the sample, and C its heat capacity. For sufficiently poor thermal conductors, the measured time constant

$$\tau_{meas} \approx \tau_2 \equiv d^2/\sqrt{90 D_c}, \quad (2)$$

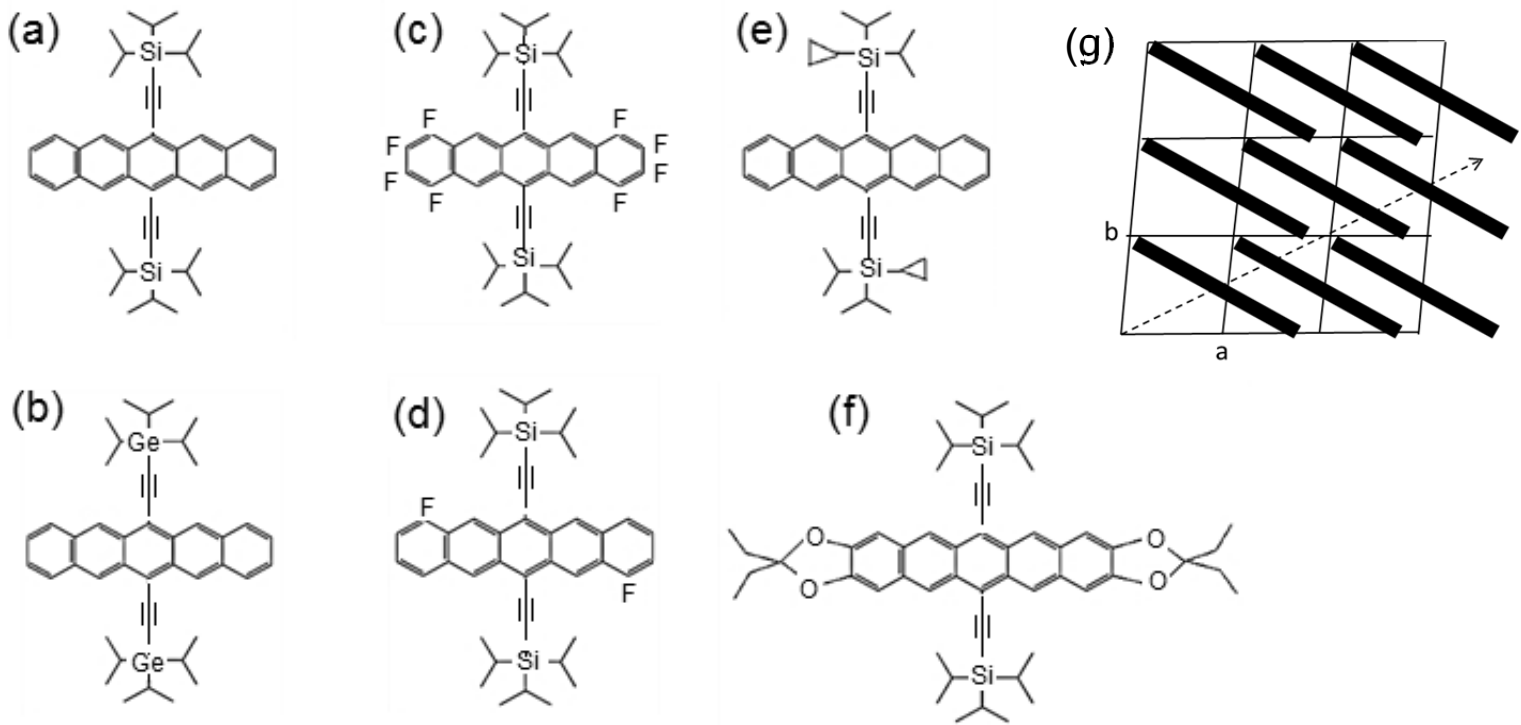

**Figure 1.** (a-f) Molecular structures of crystals studied:
(a) bis(triisopropylsilylethynyl) pentacene (TIPS-Pn [9-12]);
(b) bis(triisopropylgermylethynyl) pentacene (TIPGe-Pn) [15];
(c) bis(triisopropylsilylethynyl) octafluoropentacene ($F_8$-TIPS-Pn [13]);
(d) bis(triisopropylsilylethynyl) perifluoropentacene ($F_2$-TIPS-Pn[15]);
(e) bis(cyclopropyl-diisopropylsilylethynyl) pentacene (CP-DIPS-Pn [15]);
(f) tetraethyl-bis(triisopropylsilylethynyl)-tetraoxadicyclopenta[b,m] pentacene (EtTP-5 [14]).
(g) Schematic of the **ab**-plane structure of TIPS-Pn, with the heavy bars representing the pentacene backbones [11]. The dashed arrow shows the [2,1,0] growth axis of needle-shaped crystals.

where d is the thickness of the sample, $\tau_2$ is the internal thermal time constant describing heat flow through the thickness of the sample (i.e. along the **c**-axis for our samples) and $D_c = \kappa_c/c\rho$ is the transverse, interlayer thermal diffusivity [6, 17]. For small values of this intrinsic $\tau_2$, however, one also needs to consider the time response of the thermometer, which for our technique is dominated by the interface thermal resistance between the silver paint and sample. We've measured this interface time constant to vary between ~ 1 and 10 ms for silver paint contact areas ~ 1 mm$^2$ on different materials [6]. Because the materials studied here are slightly soluble in the butyl acetate base of the paint, we expect the interface time constants to be closer to 1 ms, but for measured time constants less than a few ms, we only conclude a lower limit for $D_c$, i.e. $D_c > d^2/\sqrt{90}\tau_{meas}$. In fact, this was the case for all the samples discussed in this paper. (In contrast, a d ≈ 90 μm crystal of rubrene had $\tau_{meas}$ ≈ 17 ms with an interface time constant < 6 ms [6].)

For longitudinal, in-plane thermal conductivity ($\kappa_{long}$) measurements, we use the technique of Hatta, *etal* [18]. A movable screen is placed between the chopped light source and sample, and the temperature oscillations are measured (with a thermocouple again glued to the sample with silver paint) on the back side of the sample in the screened portion. If $\omega\tau_2 \ll 1$, the position dependence of the oscillating temperature is given by [18]

$$d\ln T_\omega(x)/dx = -(\omega/2D_{long})^{1/2}, \quad (3)$$

where $D_{long} = \kappa_{long}/c\rho$ and x = the distance between the thermometer and edge of the screen. In our setup, the screen is attached to a micrometer (precision ± 3 μm) which measures the distance, $x_0-x$, where the offset $x_0$ depends on the relative positions of the sample and screen. For these measurements, the frequency is fixed so that the thermal response of the thermometer does not affect results. To be sure that we are in the correct frequency limit and the edge of the screen is not too close to the thermometer (i.e. overlapping with the silver paint spot) we check that we obtain the same values for the "frequency normalized" slopes, $f^{1/2} d\ln T_\omega/dx$ at different frequencies, where $f \equiv \omega/2\pi$, as shown in Figure 2. Because Eqtn. (1) assumes that both the illuminated and screened portions of the sample are much longer than the longitudinal diffusion length, $(D_{long}/\omega)^{1/2}$, which is typically ~ 0.5 mm for our samples and frequencies, data was taken on needle shaped samples ~ 1 cm long.

In Figure 2, we plot $f^{-1/2} \ln V_\omega$ as a function of position behind a screen for a TIPS-Pn crystal along its [2,1,0] needle axis at a few frequencies. From Eqtn. (3), the slope is inversely proportional to $D_{long}^{1/2}$. For large $x_0-x$, the edge of the screen overlaps the silver paint holding the thermocouple, and the signal begins to saturate, whereas for small $x_0-x$, the edge of the screen is far from the thermocouple, so $V_\omega$ is small and approaches the thermocouple offset voltage and noise level. For intermediate distances, the same slope, $f^{1/2} d \ln V_\omega /dx$, is measured for different frequencies, yielding $D_{long} = 1.0 \pm 0.1$ mm$^2$/s. (Similar values were found for other crystals.) The specific heat of TIPS-Pn, measured on a pellet with differential scanning calorimetry with a precision ~ 3% [20], is shown in the Figure 3 inset. Using c = 1.48 J/gK and density ρ = 1.1 g/cm$^3$ [7], we find $\kappa_{long}$ = 16 ± 2 mW/cm·K.

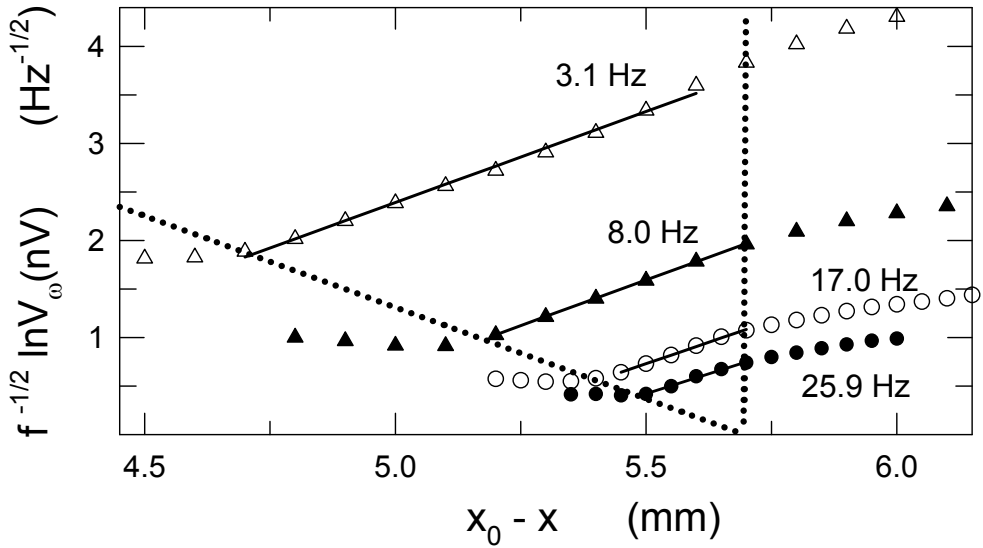

**Figure 2.** Spatial dependence of $V_\omega$ for an 8 mm long TIPS-Pn crystal at several frequencies. The region of parallel, linear variations of $f^{1/2} \ln(V_\omega)$, shown by the solid lines, is outlined by the dotted lines, as discussed in the text.

This value of $\kappa_{long}$ is several times larger than the room temperature value for rubrene [5], but comparable to the phonon thermal conductivity of quasi-one dimensional organic conductors containing segregated molecular stacks with excellent π-orbital overlap [21-23]. In fact, its value suggests that, in addition to acoustic phonons, a significant fraction of the heat is carried by low energy, propagating optical phonons. The room temperature specific heat is over thirty times the value associated with acoustic modes ($c_{acoustic}$ = 3R/M = 0.039 J/g·K, where R is the gas constant and the molecular weight M = 638 g/mole), indicating that most of the room temperature specific heat is due to excitation of optical modes. Consider

$$\kappa = (\rho/3)\Sigma c_j v_j \lambda_j, \quad (4)$$

where $c_j$, $v_j$, and $\lambda_j$ are the specific heat, propagation velocity, and mean-free path associated with phonons of mode j. If the acoustic phonon velocity is between 1 and 3 km/s (typical for molecular solids [24]) and we assume that only the acoustic modes are propagating, Eqtn. (4) implies that λ > 400 A°, i.e. > 50 in-plane lattice constants. This value is much larger than expected, since acoustic modes are expected to have significant scattering both from librations of the side groups and large thermal motion of the TIPS-Pn molecule along the long-axis of the pentacene backbone [25] (~transverse to the needle axis). If the phonon mean-free paths are in fact relatively small, much of the heat must be carried by sufficiently dispersive, low frequency ($< k_BT$) optical modes.

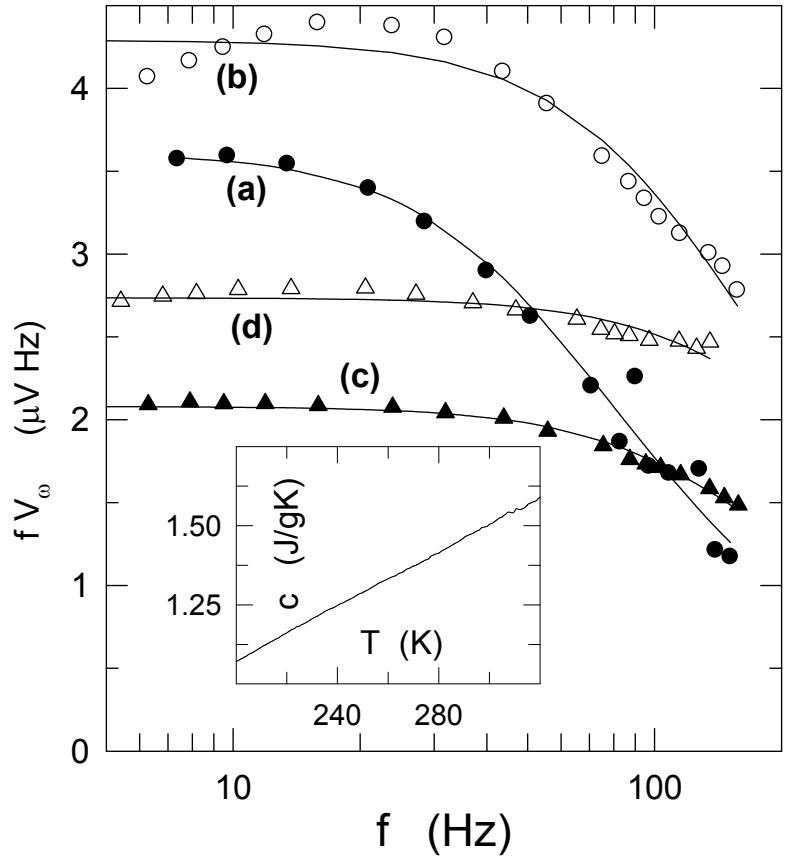

**Figure 3.** Frequency dependence of $fV_\omega$ for representative crystals; the curves show fits to Eqtn. (1b). (a) TIPS-Pn (signal x 3), d = 610 μm, $\tau_{meas}$ = 2.8 ms → $D_c$ > 14 mm$^2$/s;
(b) F$_2$-TIPS-Pn, d ≈ 300 μm, $\tau_{meas}$ = 1.26 ms → $D_c$ > 7 mm$^2$/s;
(c) TIPGe-Pn (signal x 2), d = 460 μm, $\tau_{meas}$ = 1.03 ms → $D_c$ > 20 mm$^2$/s;
(d) EtTP-5, d ≈ 300 μm, $\tau_{meas}$ = 0.68 ms → $D_c$ > 14 mm$^2$/s.
Inset: Specific heat of TIPS-Pn.

Figure 3 shows the frequency dependence of $fV_\omega$ of a d=610 μm TIPS-Pn crystal, with $\tau_{meas}$ = 2.8 ms. In Ref. [6], we showed results for a d=335 μm crystal, with $\tau_{meas}$ = 1.3 ms; much thinner crystals, e.g. d ≈ 100 μm, had similar time constants. These values of $\tau_{meas}$ and their lack of correlation with thickness imply that we are limited by the thermal interface resistance of the silver paint, and from the 610 μm crystal (the thickest available), we deduce $D_c$ > 14 mm$^2$/s and $\kappa_c$ > 225 mW/cm$^2$·K .

These values are much larger than typically found in a van-der-Waals bonded molecular crystal (e.g. for rubrene we measured $D_c$ ~ 0.05 mm$^2$/s, giving $\kappa_c$ ~ 0.7 [6], while for pentacene $\kappa_c$ = 5.1 mW/cm$^2$·K [26]), and are comparable to the values for materials with extended bonding (e.g. Al$_2$O$_3$ has D ~ 10 mm$^2$/s and κ ~ 300 mW/cm·K [27]). Also surprising is the anisotropy, $\kappa_c$ > 14 $\kappa_{long}$; because the in-plane (π-π) interactions were assumed to be stronger than the inter-plane (hydrocarbon) interactions, we expected the longitudinal thermal conductivity to be greater than the transverse, as we observed for rubrene, $\kappa_c$ ~ $\kappa_{long}$/6 [6]. If one assumes that thermal transport is dominated by intermolecular thermal resistances and that these were *equal* for **c**-axis and in-plane interactions, then the thermal conductivities in different directions would be proportional to the packing densities in their transverse planes [28], making $\kappa_c$ ~ 2 $\kappa_{long}$ because a ~ b ~ c/2 [9]. The much larger anisotropy we measure therefore indicates that the intermolecular thermal resistances along **c** are *smaller* than those in the plane, implying stronger phonon interactions between the TIPS side groups than between the in-plane pentacene backbones. For example, the isopropyl groups of molecules in neighboring layers are only separated by ~ 0.4 nm [11]. Because of the slight ionicity of the C-H bond, there will be significant Coulomb coupling of vibrations (e.g. librations) in these groups, increasing the **c**-axis dispersion of these low-frequency phonons. (If instead, one assumes that most of the heat is carried by acoustic phonons, which seems to be the case for rubrene [6], then the same analysis used for $\kappa_{long}$ would give an unreasonably large interlayer phonon mean-free-path $\lambda_c$ > 300c!)

To check these results for TIPS-Pn, we measured the transverse and longitudinal diffusivities of the related materials listed in Figure 1. Figure 3 shows the results of the frequency dependence of $fV_\omega$ for some crystals, along with the values of d and $\tau_{meas}$. F$_2$-TIPS-Pn crystals are needles with a brick-layer structure [15] similar to that of TIPS-Pn (Figure 1g). TIPGe-Pn and EtTP-5 crystals, on the other hand, grow as thick (d > 200 μm) flakes (in-plane dimensions < 3 mm). Molecules in EtTP-5 are coplanar but insulating substituents keep the aromatic faces ~ 10 A$^o$ apart along [010] [14]. In TIPGe-Pn, the orientation of the pentacene backbones alternate in the **ab**-plane so that the aromatic surface of each molecule faces insulating substituents of adjacent molecules [15]. In both materials, therefore, there is poor π-orbital overlap in the **ab**-plane, but the silyl and germyl side groups still extend along the interlayer **c**-axis [14,15]. For all materials, a few crystals were measured with representative results shown. In all cases, the responses are very fast so only lower limits for $D_c$ (given in the caption) were determined. As for TIPS-Pn, the large diffusivities are unusual for van-der-Waals bonded molecular crystals and suggest strong phonon interactions between the side groups, giving significant dispersion to low-frequency optical phonons which consequently carry most of the heat.

Figure 4 shows representative spatial dependences of $f^{-1/2}\ln(V_\omega)$ for needle-shaped crystals of $F_2$-TIPS-Pn, $F_8$-TIPS-Pn, and CP-DIPS-Pn; for the latter two, all crystals were very thin (< 100 µm) so frequency-dependent transverse measurements are not meaningful. All of these have brick-layer structures similar to that of TIPS-Pn [13-15]. Because the cyclopropyl group makes the side groups slightly more rigid in CP-DIPS-Pn than in the other crystals, the molecules are more tightly packed; e.g. the unit cell of CP-DIPS-Pn is 6% smaller than that of TIPS-Pn [15]. The measured slopes and calculated longitudinal diffusivities for the materials are given in the caption. As for TIPS-Pn, slopes were measured at a few frequencies for each crystal and the uncertainties given in the table reflect the variations in slopes. The fluorinated crystals may have slightly steeper slopes, and therefore lower thermal diffusivities, than TIPS-Pn, but the differences are less than the uncertainties in the measurement. The slopes for CP-DIPS-Pn had more scatter, probably reflecting the somewhat shorter crystals available, but were always considerably steeper than for the other compounds. The resulting lower diffusivity for CP-DIPS-Pn is surprising because one would expect the more rigid side-groups to reduce scattering of acoustic phonons. It again suggests that much of the heat is carried by molecular vibrations and that the greater rigidity of the side-groups reduces the dispersion of the relevant low-energy modes.

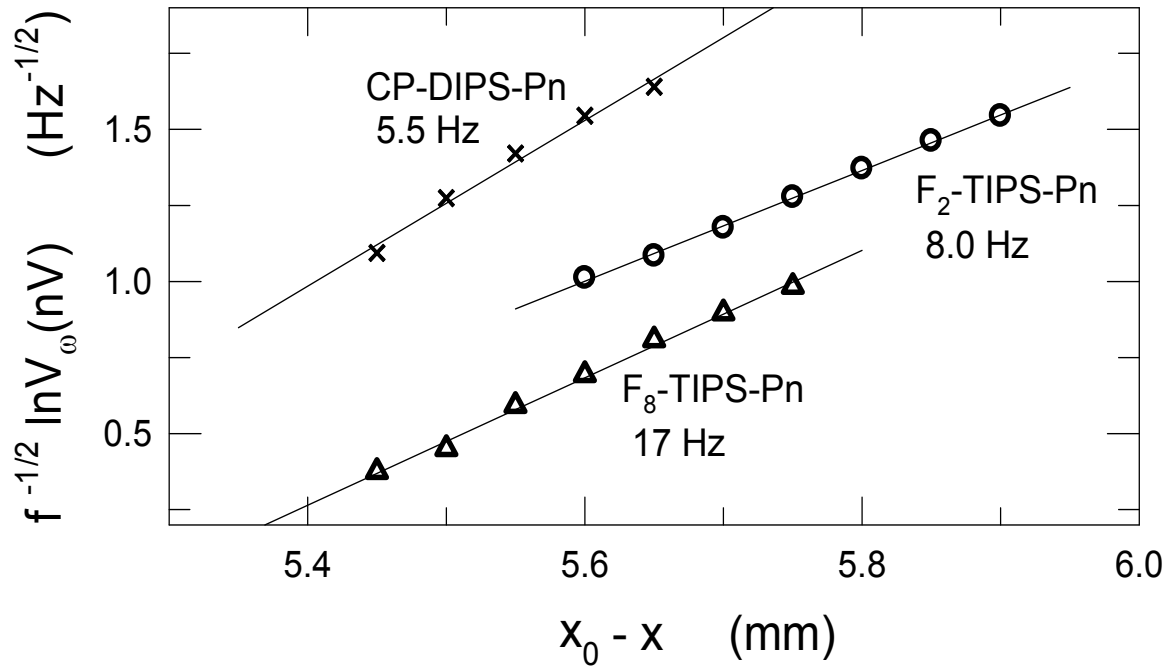

**Figure 4.** Spatial dependence of $V_\omega$ for crystals of CP-DIPS-Pn (length = 6 mm, $-f^{-1/2}\,d\ln V_\omega/dx = 2.7 \pm 0.3$ $(Hz^{1/2}\cdot mm)^{-1}$, $D_{long} = 0.43 \pm 0.10$ mm²/s), $F_2$-TIPS-Pn (length = 10 mm, $-f^{-1/2}\,d\ln V_\omega/dx = 1.89 \pm 0.07$ $(Hz^{1/2}\cdot mm)^{-1}$, $D_{long} = 0.88 \pm 0.09$ mm²/s), and $F_8$-TIPS-Pn (length = 11 mm, $-f^{-1/2}\,d\ln V_\omega/dx = 1.98 \pm 0.10$ $(Hz^{1/2}\cdot mm)^{-1}$, $D_{long} = 0.80 \pm 0.10$ mm²/s) at selected frequencies in their linear regions.

The large thermal diffusivities indicate that, for those of these materials with brick-layer structures and high electronic mobility, Joule heating of micro-electronic components should not pose a problem. While the longitudinal diffusivities are slightly larger than desired for thermoelectric applications, they are not excessive. However, thermoelectric devices would need to be constructed so that the very high transverse diffusivities do not create thermal shunts.

In summary, we have measured the longitudinal (needle-axis) and transverse, interlayer thermal conductivities of crystals of TIPS-Pn by ac-calorimetry. We have found that the longitudinal value is higher than that of rubrene [5] and pentacene [26] and comparable to quasi-one dimensional conductors with excellent π-orbital overlap [21-23]. The transverse thermal diffusivity is at least an order of magnitude *larger* than the longitudinal. These values and inverted anisotropy of κ indicate that molecular vibrations, presumably concentrated on the silyl-containing side groups, have sufficient intermolecular interactions and dispersion to carry most of the heat. Similar values for both the in-plane and interlayer thermal diffusivities were found for several other materials with related structures.

This research was supported in part by the National Science Foundation, grant DMR-1262261. The synthesis and crystal engineering of organic semiconductors was supported by the Office of Naval Research, grant N00014-11-1-0328.